\documentstyle[12pt]{article}
\advance\textheight by \topskip
\newcommand{\BE}{\begin{equation}}
\newcommand{\EE}{\end{equation}}
\newcommand{\BA}{\begin{eqnarray}}
\newcommand{\EA}{\end{eqnarray}}

\def\xt{\tilde{x}}
\def\Rt{\tilde{R}}
\def\e{\varepsilon}
\def\s{\sigma}
\def\BH{\beta_{H}}
\def\ad{\alpha ' }
\def\B{\beta}
\def\p{\partial}

\def\mR{\frac{m}{R}}

\def\Nt{\tilde{N}}
\def\ti{\tau_{1}}
\def\tii{\tau_{2}}

\def\Rb{\bar{R}}
\def\ET{e^{2 \pi i \tau }}
\def\et{e^{4 \pi \tii }}
\def\pa{2 \pi \ad }
\def\pt{2 \pi \tii }

\def\mw{{\displaystyle \left\langle \left( \sum_{i=1}^{25} ( m_{i} w_{i} )^{2} \right) \right\rangle}}
\def\bmw{\langle ( \sum_{i=1}^{25} ( m_{i} w_{i} )^{2} ) \rangle}

\def\one{\langle 1 \rangle}
\def\f{\mid f( \ET ) \mid}
\def\T{\sum_{r=1}^{ \infty } \exp \left( - \frac{r^{2} \beta^{2}}{4 \pi \ad \tii } \right)}
\def\TrB{\sum_{r=1}^{ \infty } (-r \beta) \exp \left( - \frac{r^{2} \beta^{2}}{4 \pi \ad \tii } \right)}
\def\O{\Omega}
\def\no{\nonumber}
\def\L{\Lambda}
\def\FB{F_{B}}

\def\bi{\bibitem}
\def\TS{\sum_{r=1}^{ \infty } (1-(-1)^{r}) \exp \left( - \frac{r^{2} \beta^{2}}{4 \pi \ad \tii } \right)}

\def\FS{F_{S}}
\def\thiv{\mid \vartheta_{4} (0 \mid 2 \tau ) \mid^{-16}}
\def\mws{{\displaystyle \left\langle \left( \sum_{i=1}^{9} ( m_{i} w_{i} )^{2} \right) \right\rangle}}
\def\smw{\langle ( \sum_{i=1}^{9} ( m_{i} w_{i} )^{2} ) \rangle}
\def\db{\bar{d}}
\def\l{\lambda}
\def\vp{\varphi}
\def\G{\Gamma}

\input epsf

\begin{document}

\rightline{May 97}
\rightline{OU-HET 265}
\rightline{hep-th/9705099}

\vspace{.8cm}
\begin{center}
{\large\bf Correlation between Momentum Modes and Winding Modes in Brandenberger-Vafa's String Cosmological Model}

\vskip .9 cm

{\bf Kenji Hotta,}
\footnote{E-mail address: hotta@funpth.phys.sci.osaka-u.ac.jp}
{\bf Keiji Kikkawa,}
\footnote{E-mail address: kikkawa@funpth.phys.sci.osaka-u.ac.jp}

Department of Physics, Osaka University, \break
Toyonaka, Osaka 560, Japan \break 

 \vskip -1ex
    and \break
 \vskip -1ex

{\bf Hiroshi Kunitomo}
\footnote{E-mail address: kunitomo@yukawa.kyoto-u.ac.jp}

Yukawa Institute for Theoretical Physics, Kyoto University, \break
Sakyo-ku, Kyoto 606-01, Japan

\vskip.6cm

\end{center}

\begin{quotation}

Brandenberger and Vafa have proposed the string cosmological model based on T-duality. In this model, they took the toroidal target space and introduced 
the new position coordinate $\tilde{x}$ conjugate to the winding mode, in addition to the position $x$ conjugate to the momentum mode. In this way they can describe a universe larger than the string scale with the coordinate $x$ and the one smaller with the coordinate $\xt$ . Resultingly, they never encounter the singularity seen in the standard Big Bang scenario. The most interesting phenomenon in this model is the transition from $\xt$-space to $x$-space when the size of universe is nearly the string scale. Here, we define the dispersion of the momentum number $m$ times the winding number $w$ as the `correlation' of momentum modes and winding modes. Then using the statistical mechanics of strings on a torus, we calculate the correlation in low and high temperature limits, and we study the possibility that we can observe this effect today, but we will see that this is unlikely.

\end{quotation}

\normalsize
\newpage

\section{Introduction}

One of the most interesting open questions in theoretical physics is the initial singularity problem in the early universe. In particular, Penrose and Hawking's singularity theorem has made it more serious. If string theory is the ``theory of everything", it should be able to solve this question. Recently, Brandenberger and Vafa \cite{BVcos} proposed a string cosmological model which may avoid the initial singularity due to the T-duality. \cite{Kik} There, when one direction of the target space is compactified into $S^{1}$, this duality implies the invariance of physical law by transforming the radius of $S^{1}$ into its inverse and exchanging the quantum number of momentum modes for that of winding modes. Therefore, in addition to the coordinate $x$ conjugate to the momentum mode, Brandenberger and Vafa introduce the dual coordinate $\xt$ conjugate to the winding mode. Then, we can realize that the universe starts from a small size $x$-space, which is large in $\xt$-space, to a large size $x$-space. In this model, therefore, it is possible that entire universe has never been shrunken to a point-like singular space.

On the other hand, string theory has no direct experimental evidence, since we cannot obtain string scale energy with present experimental equipment. Therefore, it is natural to investigate cosmological observations in order to look for evidence that string theory is valid. In this paper, we study the `correlations' between momentum modes and winding modes, which are defined in the next section, and the possibility that we can observe a trace of the $\xt$ universe today.

The plan of this paper is as follows: In this section, we review the Brandenberger-Vafa model and the T-duality. Then in \S \ref{sec:correlation} we define the `correlations' of momentum modes and winding modes and describe the motivation of our work. In \S \S \ref{sec:low} and \ref{sec:high}, we give the calculation of the momentum-winding correlations in low and high temperature limits, and then we discuss the possibility of observing the trace of the $\xt$-space in \S \ref{sec:inflation}. Finally, summary is presented in \S \ref{sec:conclusion}.

Before we describe the Brandenberger-Vafa model, we first review T-duality. \cite{Tdual} As the simplest example of this duality, we consider a single bosonic string coordinate compactified on a circle of radius $R$. If we use the orthogonal gauge for the world sheet metric, then we can write the mass spectrum as
\BE
  M^{2} = \frac{\ad}{2}  \left( \mR \right)^{2} + \frac{1}{2 \ad} (wR)^{2} + N + \Nt ,
\label{eq:1Dmass}
\EE
where the first term represents the energy of momentum modes, the second the winding modes and the third and forth terms oscillation modes. If we define the dimensionless radius $\Rb$ as
\BE
  \Rb = \frac{R}{ \sqrt{ \ad }} ,
\EE
this mass spectrum is invariant under the duality transformation
\BA
  \Rb \longleftrightarrow \frac{1}{\Rb}, \no \\
  m \longleftrightarrow w .
\label{eq:Tdualtrans}
\EA
Not only mass spectrum, but the entire string theory is invariant under this duality transformation, and this is the T-duality.

As mentioned above, Brandenberger and Vafa have defined a new coordinate based on this duality as follows. \cite{BVcos} In general relativity, we can define the distance between two positions by measuring the time which a light ray takes from one position to the other and dividing it by the light velocity. If we wish to know the length precisely, we must compose many momentum modes and make sharply localized wave packets. In ordinary quantum field theory, we must define the position to be the Fourier transform of momentum states:
\BE
\vert x \rangle = \sum_{p}\exp (ix \cdot p) \vert p \rangle . \no
\EE
In the toroidal universe we are considering, it is difficult to make the wave packets when $\Rb$ is much smaller than $1$, since it is extremely energetic, as determined by the uncertainty relation
\BE
\triangle x \cdot \triangle p \geq \hbar . \no
\EE
In string theory, however, we can introduce a new position coordinate as the Fourier transform of winding number states:
\BE
\vert \xt \rangle = \sum_{w}\exp (i \xt \cdot w) \vert w \rangle .
\EE
It is energetically easier to form the wave packets of winding modes than those of momentum modes. Since we are considering a torus target space, the periodicity of the coordinates is
\BA
  \vert x \rangle &=&  \left\vert x + 2 \pi \Rb \sqrt{ \ad} \right\rangle \no \\
  \vert \xt \rangle
     &=&  \left\vert \xt + \frac{2 \pi}{\Rb} \sqrt{ \ad} \right\rangle
     =  \left\vert \xt + 2 \pi \Rt \sqrt{ \ad} \right\rangle , \no
\EA
where we define the inverse radius $\Rt$ by
\BE
  \Rt = \frac{1}{\Rb} ,
\EE
which is the radius of the universe in $\xt$ coordinates.
\begin{figure}[ht]
\begin{center}
$${\epsfxsize=9.5 truecm \epsfbox{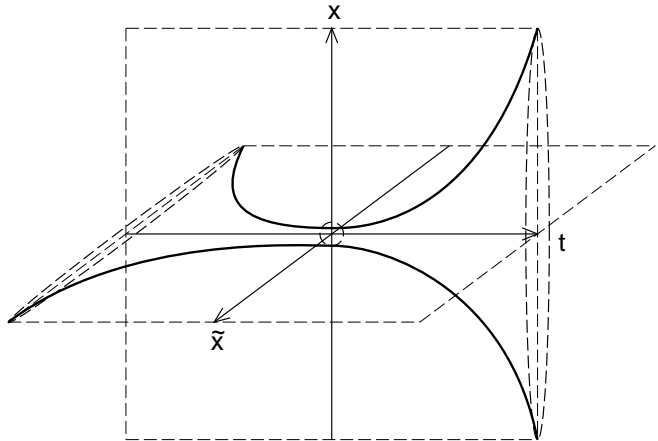}}$$
\caption{The time development of the scale factors in the Brandenberger-Vafa model.}
\label{fig:x_xt_cosmo}
\end{center}
\end{figure}

Now, let us explain the cosmological scenario. In the present, the universe is expanding and we can take the coordinate of the universe as $x$. But, as one traces backward in time, the universe shrinks, and if $\Rb$ is smaller than $1$ we describe it with $\xt$, and the dual radius $\Rt$ becomes larger and larger. To put it another way, at the beginning, the universe is very large in $\xt$ and shrinks to $\Rb = 1$. Then the universe enters into $x$-space, and the radius becomes larger as time goes on. In Fig.\ref{fig:x_xt_cosmo}, we represent the situation in which the universe expands in the coordinate $x$ monotonically and shrinks in $\xt$. If we suppose that the radius does not diverge in both coordinates, there is no singularity. As Brandenberger and Vafa mentioned, the transitions between $\xt$- and $x$-spaces could occur many times, i.e., an oscillating universe. In any case, the most interesting physical phenomena can take place when this transition happens. In the next section, we investigate this region.

\section{Correlation between the two modes}
\label{sec:correlation}

As mentioned in the previous section, according to the Brandenberger-Vafa model, the universe shrinks in the coordinate $\xt$ first, and the radius of universe becomes a minimum at a certain point ($\Rb = 1$). Then it expands in the coordinate $x$. The important point to note here is that the way of measuring the size of the universe is switched from $\xt$ to $x$, where $\xt$ and $x$ are conjugate to the winding number $w$ and momentum $m$, respectively. Around the transition area ($\Rb \sim 1$) `the correlation' of modes is defined as follows. In the $\Rb < 1$ region, the winding modes dominate, and wave packets are made of superpositions of winding modes, while in $\Rb > 1$ region the wave packet, are mostly made of momentum modes. As $\Rb$ grows from the $\Rb < 1$ to the $\Rb > 1$ region, the two modes must be correlated. As a representation of this `correlation', we choose the dispersion of $m$ times $w$,
\BE
{ \triangle (mw)}^2 \equiv \langle (mw)^2 \rangle - ( \langle mw \rangle )^2 ,
\EE
where $\langle \ \ \ \rangle$ denotes the statistical average for a single string. In our calculation, the second term on the RHS, $\langle mw \rangle$ is vanishing since both $m$ and $w$ can assume positive and negative values, so that ${ \triangle (mw)}^2 \equiv \langle (mw)^2 \rangle$. We explain why this dispersion represents `the correlation'.
\begin{figure}[ht]
\begin{center}
$${\epsfxsize=11.5 truecm \epsfbox{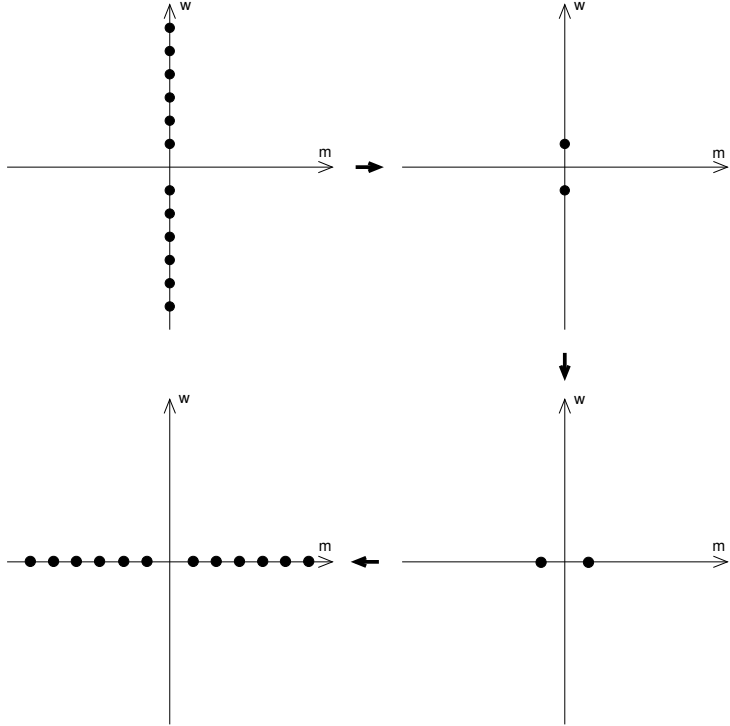}}$$
\caption{The quantum transition of momentum modes and winding modes at low temperature.}
\label{fig:mw_transition_L}
\end{center}
\end{figure}
\begin{figure}[ht]
\begin{center}
$${\epsfxsize=11.5 truecm \epsfbox{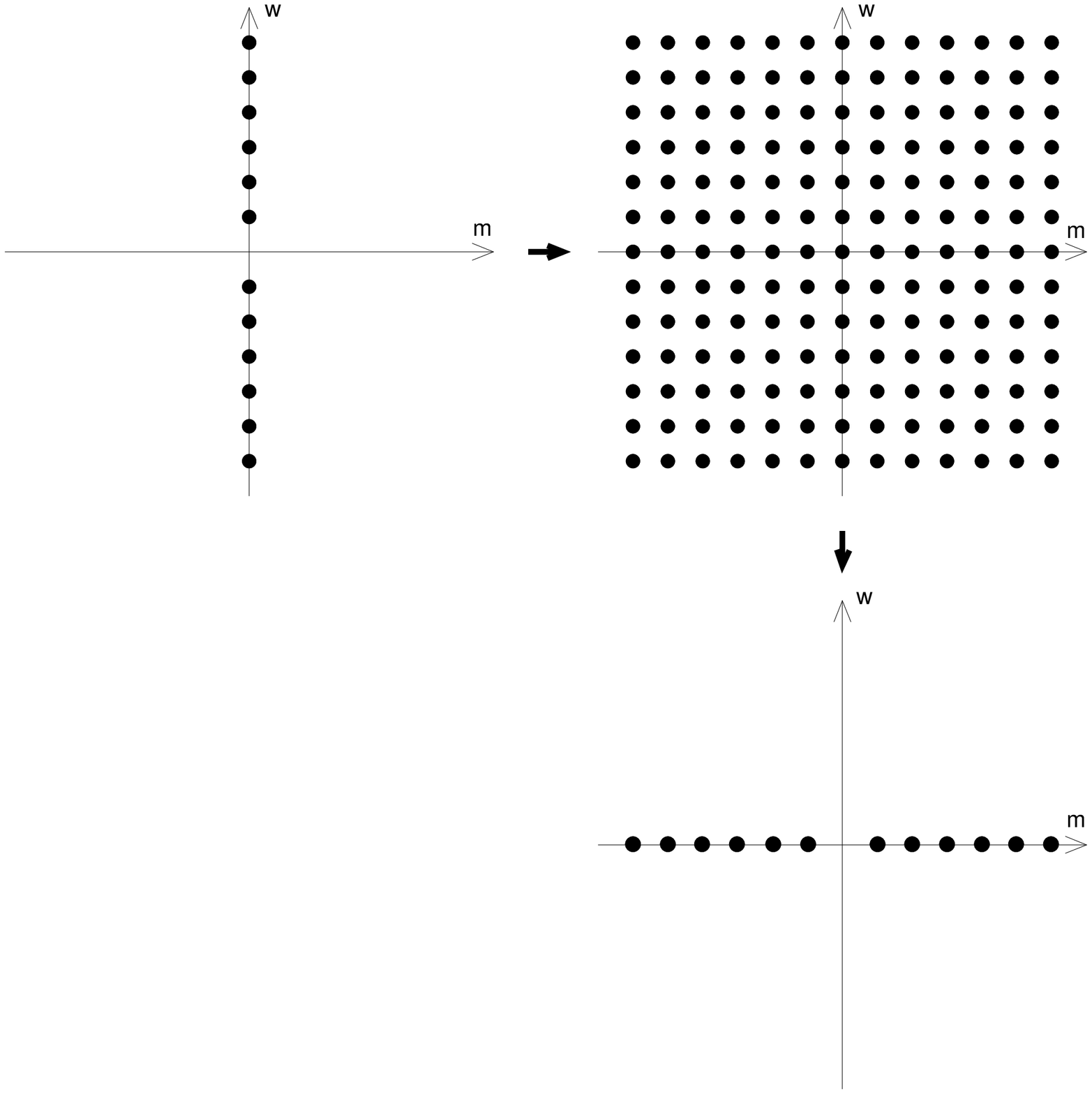}}$$
\caption{The quantum transition of momentum modes and winding modes at high temperature.}
\label{fig:mw_transition_H}
\end{center}
\end{figure}

Assume that there is a wave packet in the $x$-space with extension $\triangle x$, and in the $\xt$-space with extension $\triangle \xt$. From uncertainty principle in each space, 
\BA
{ \displaystyle
  \triangle x \cdot \frac{ \triangle m }{{\ad}^{1/2} \Rb} \geq \hbar, } \no \\
{ \displaystyle
  \triangle \xt \cdot \frac{\triangle w \Rb}{{\ad}^{1/2}} \geq \hbar }. \no
\EA
These two inequalities are combined into 
\BE
 \triangle x \cdot \triangle \xt \geq \frac{ \hbar^2 }
      { \triangle m \cdot \triangle w } \ad.
\label{eq:xxuncertainty}
\EE
Since the magnitude of $\triangle m \cdot \triangle w$ is related to that of $\triangle (mw)$, as will be discussed in the following sections, we can determine that the possibility of making sharp wave packets in both $x$-space and $\xt$-space. From this point of view, we simply call the dispersion $\triangle (mw)$ `the momentum-winding correlation'. Moreover, if this `momentum-winding correlation' is very large, we assume that the wave packet in $\xt$-(or $x$-)space can carry information to the wave packet in $x$-(or $\xt$-) space. In other words, the information in the world with $\Rb < 1$ stocked in the wave packets in $\xt$-space can be conveyed to an observer in $x$-space where $\Rb > 1$ if the correlation is large.

Next, let us first consider the low temperature case. When $\Rb \ll 1$, string states are dominated by winding modes, and as $\Rb$ grows, low energy winding modes survive, and eventually only $w \sim 0$ modes remain when $\Rb \sim 1$. These states may be converted into small momentum modes $m \sim 0$, and then gradually be dominated larger momentum modes as $\Rb$ increases. Therefore, in the low temperature limit, there can be strings with $m \sim 0$ and $w \sim 0$ modes. The number of modes are so restricted in this case that no communication of information is possible.

On the other hand, in the high temperature limit around $\Rb \sim 1$, there can be many modes with $m \neq 0$ and $w \neq 0$. If the number of such modes is sufficiently large, we can convey information with these modes from the $\xt$ universe to the $x$ universe. Figures \ref{fig:mw_transition_L} and \ref{fig:mw_transition_H} indicate these permitted modes by dots at $m$-$w$ plane at low and high temperature. In the low temperature limit, almost no modes are excited as $\Rb$ approaches $1$, while at high temperature, many modes are excited in both winding and momentum modes. We discuss these correlations in the low and high temperature limit in the next two sections.

\section{The momentum-winding correlation in low \\ temperature limit}
\label{sec:low}

\begin{figure}
\begin{center}
$${\epsfxsize=9.5 truecm \epsfbox{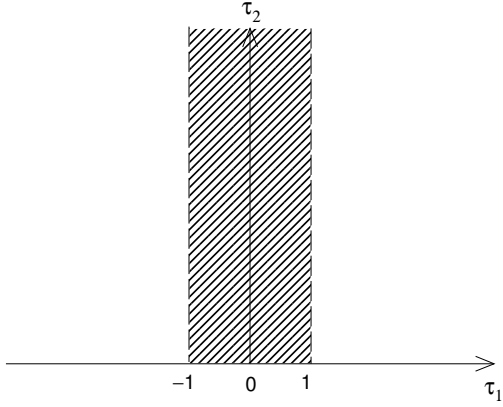}}$$
\caption{E region.}
\label{fig:E_rep}
\end{center}
\end{figure}

Before we show the momentum-winding correlations in low temperature limit, we consider an ideal gas of bosonic strings in 26 dimensions and describe how to compute the statistical average for it. In Matsubara formalism, the free energy of a string gas can be evaluated from the path integral of a connected graph of strings on the space where Euclidean time direction is compactified with the period of inverse temperature $\B$. In the ideal gas limit, we assume that there is no interaction, so we take into account only  torus world sheets which wind the Euclidean time direction at least once. We can express it as \cite{Pol},\footnote{Of course, we can rewrite this free energy in modular invariant form --- a dual representation in which the integration region is chosen in a fundamental region. \cite{Tan1} But above we use the E-representation \cite{Tan1} for simplicity of calculation.}
\BA
  \FB ( \B ) = - \frac{V}{ ( \pa )^{13} } \int_{E} 
    \frac{d^{2} \tau }{ ( \pt )^{2} } ( \pt )^{-12} \et 
    ( \f^{-48} - 1 ) \no \\
    \times \left[ \T \right],
\label{eq:bosonfreenoncom}
\EA
where $\tau = \ti + i \tii$, $\B$ is the inverse temperature, $V$ is the volume that we are considering, and $f(z)$ is
\BE
  f(z) = \prod_{n=1}^{ \infty } ( 1 - z^{n} ). \no
\EE
Here, in order to remove the tachyon modes of the bosonic string, we use the replacement $\f^{-48} \ \ \ \rightarrow \ \ \ \f^{-48} - 1$, and the integral region is shown in the region $E$ indicated in Fig.\ref{fig:E_rep}. We can also evaluate the free energy on the 25-dimensional torus $T^{25}$ whose radius is $\Rb$: \cite{Ope}
\BA
  \FB ( \B ) = - \frac{1}{ ( \pa )^{\frac{1}{2}} } \int_{E} 
    \frac{d^{2} \tau }{ ( \pt )^{2} } ( \pt )^{-12} \et 
    ( \f^{-48} - 1 ) \no \\
    \times  (C_{1})^{25} \left[ \T \right],
\label{eq:comfree}
\EA
where $C_{1}$ is
\BE
  C_{1} ( \Rb ,  \tau ) 
   = \sqrt{ \pt } \sum_{m,w} \exp \left[ 2 \pi i \ti mw
      - \pi \tii \left( \frac{m^{2}}{\Rb} + w^{2} \Rb^{2} \right) \right].
\label{eq:Ci}
\EE
From (\ref{eq:Ci}), we find that the free energy (\ref{eq:comfree}) is invariant under the duality transformation (\ref{eq:Tdualtrans}), as expected. In this form, however, it is difficult to understand how to evaluate the statistical averages about the string gas. Therefore, we reconsider these from the foundations of statistical mechanics. 

In the canonical ensemble method, the partition function of an ideal gas can be written as
\BE
  Z = \sum_{n_{\s} = 1}^{\infty} 
        \exp \left[ - \beta \sum_{\s} \e_{\s} n_{\s} \right],
\label{eq:partition}
\EE
where $\e_{\s}$ denotes the energy of a single string in a state $\s$, and $n_{\s}$ the number of strings in it. Since we are considering the on-shell bosonic string gas, the energy of a single string $p_{0}$ in the 26-dimensional flat (noncompact) space-time is given by
\BA
  p_{0} &=& \sqrt{ \vec{p}^{\ 2} + M^{2}} \no \\
        &=& \sqrt{ \sum_{i=1}^{25} (p_{i})^{2} + \frac{2}{\ad} (N + \Nt - 2) },
\EA
where the string mass $M$ is
\BE
  M^{2} = \frac{2}{\ad} (N + \Nt - 2).
\EE
There is the constraint equation
\BE
  N - \Nt = 0. \no
\label{eq:NNT}
\EE
The free energy of the string gas is given by
\BA
  \FB ( \B ) &=& \frac{V}{ \B } \sum_{M^{2}}
       \int \frac{d^{\ 25} p}{(2 \pi )^{25}} \ln (1- e^{- \beta p_{0}}) \no \\
    &=& - \sum_{M^{2}} \frac{1}{ \beta } \sum_{r=1}^{\infty}
     \sum_{n=0}^{\infty} 
      \frac{1}{r} \frac{(-r \beta )^{n}}{n!} 
         \int \frac{d^{\ 25} p}{(2 \pi )^{25}} ( \vec{p}^{\ 2} + M^{2} )^{n/2},
\label{eq:massfree}
\EA
where we have expanded logarithmic and exponential functions in the second equality. Using the formula with the Gamma function
\BE
  \omega^{- \alpha } = \frac{1}{2^{ \alpha } \G ( \alpha )} \int_{0}^{\infty}
      \exp \left(- \frac{\omega s}{2} \right) s^{\alpha -1} ds,
\EE
we can integrate $p$ out
\BE
  \int \frac{d^{\ 25} p}{(2 \pi )^{25}} ( \vec{p}^{\ 2} + M^{2} )^{n/2}
    = \frac{(2 \pi )^{1/2}}{2^{-n/2} \G (- \frac{n}{2} ) }
      \int_{0}^{\infty} \frac{ds}{s} (2 \pi s)^{-13} s^{ \frac{n-1}{2} }
      e^{ - \frac{1}{2} M^{2} s }.
\EE
If we take into consideration the formulae
\BA
  \G (-l) &=& \pm \infty, \\
  \G \left( -l + \frac{1}{2} \right) &=& \frac{(-4)^{l} l!}{(2l)!} \sqrt{\pi},
     \ \ \ \ \ \ \ l = 0,1,2, \cdot \cdot \cdot
\EA
the free energy is
\BE
   \FB ( \B ) = - V \sum_{M^{2}} \int_{0}^{\infty} \frac{ds}{s} (2 \pi s)^{-13}
      \sum_{r=1}^{\infty} \exp \left( - \frac{1}{2} M^{2} s 
        - \frac{r^{2} \beta^{2}}{2s} \right).
\EE
This is the free energy in proper time form. If we take account of the constraint (\ref{eq:NNT}) and perform the variable transformation
\BE
  s = \pa \tii,
\EE
we then obtain the free energy (\ref{eq:bosonfreenoncom}). Similarly, we can calculate the free energy on $T^{25}$ by using the energy of a single string,
\BE
  p_{0} = \sqrt{ \sum_{i=1}^{25} \frac{1}{{\ad}^{1/2}}
   \left[ \left( \frac{m_{i}}{\Rb} \right)^{2} + (w_{i} \Rb)^{2} \right]
      + \frac{2}{\ad} (N + \Nt - 2) }, \no
\EE
and the constraint equation
\BE
  N - \Nt = \sum_{i=1}^{25} m_{i} w_{i}, \no
\EE
and we obtain the free energy (\ref{eq:comfree}).

Next, we introduce the external field $J$, and calculate the average of some physical quantity $A$. To do this, we replace $\e_{\s}$ in (\ref{eq:partition}) by $\e_{\s} + JA$, and take the derivative with respect to $J$. Then,
\BE
 \langle A \rangle 
   = \frac{ - \frac{1}{\beta}
         \frac{ \p }{\p J} Z \mid_{J=0}}{Z \mid_{J=0}}
   = \left. \frac{ \p }{ \p J} F( \beta , J ) \right|_{J=0}.
\label{eq:average}
\EE
The result is
\BA
 \langle A \rangle 
   \!\!\! &=& \!\!\! - \left. \frac{V}{ \beta }
         \sum_{M^{2}} \sum_{r=1}^{\infty}
         \sum_{n=0}^{\infty} \frac{1}{r} \frac{(-r \beta )^{n}}{n!} 
         \int \frac{d^{\ 25} p}{(2 \pi )^{25}} 
         \frac{ \p }{ \p J} [( \vec{p}^{\ 2} + M^{2} )^{1/2} + JA]^{n}
         \right|_{J=0} \no \\
   \!\!\! &=& \!\!\! - \frac{V}{ \beta } \sum_{M^{2}}
         \sum_{r=1}^{\infty} \sum_{n=0}^{\infty} 
         (-r \beta A) \frac{1}{r} \frac{(-r \beta )^{n}}{n!} 
         \int \frac{d^{\ 25} p}{(2 \pi )^{25}} 
         ( \vec{p}^{\ 2} + M^{2} )^{n/2},
\label{eq:averageA}
\EA
where we have replaced $n-1$ with $n$ in the second equality. The difference between this formula and the free energy (\ref{eq:massfree}) is only the factor $(-r \beta A)$. For the case $A = \bmw$, taking into account the fact that the factor \( (C_{1})^{25} \) in the free energy (\ref{eq:comfree}) is replaced by
\BA
  &(\pt)^{25/2}&
   \!\!\! \sum_{m,w} \left( \sum_{i=1}^{25} m_{i}^{2} w_{i}^{2} \right) 
    \prod_{j=1}^{25} \exp \left[ 2 \pi i \ti m_{j} w_{j} - \pi \tii
     \left( \frac{m_{j}^{2}}{\Rb} + w_{j}^{2} \Rb^{2} \right) \right] \no \\
    &=& \!\!\!\!\!\!\!\!\!\!\!\! (C_{1})^{24} \frac{25}{(2 \pi i)^{2}}
      \frac{\p^{2}}{\p \ti^{2}}(C_{1}),
\EA
we can obtain
\BA
  \mw
   = - \frac{1}{( \pa )^{\frac{1}{2}}}
   \int_{E} \frac{d^{2} \tau }{ ( \pt )^{2} } ( \pt )^{-12} \et 
   ( \f^{-48} - 1 ) \no \\
   \times \left[ \TrB \right]
   (C_{1})^{24} \frac{25}{(2 \pi i)^{2}} 
   \frac{\p^{2}}{\p \ti^{2}}(C_{1}).
\label{eq:average,mw}
\EA
We must note that this average is taken over all strings of the system, so if we wish to take the average about one string, we must divide it by the number of strings in the system. This corresponds to the average of $A=1$, i.e.,
\BA
  \one = - \frac{1}{ ( \pa )^{\frac{1}{2}} } \int_{0}^{\infty} 
    \frac{d \tii }{ \pt } \int_{-1/2}^{1/2} d \ti ( \pt )^{-13} \et 
    \f^{-48} \no \\
   \times (C_{1})^{25} \left[ \sum_{r=1}^{ \infty } (-r \beta) 
   \exp \left( - \frac{r^{2} \beta^{2}}{4 \pi \ad \tii } \right) \right].
\EA
Therefore, $\bmw / \one$ is what we want to know as the momentum-winding correlation.

Now let us compute the momentum-winding correlation in the low temperature limit. If we take the large $\B$ limit in (\ref{eq:average,mw}), the part which concerns the temperature consists of the terms in the square brackets, so that this integrand contributes only when $\tii$ is very large due to the exponential factor. In the large $\tii$ limit, we can expand the parts of integrand as follows:
\BA
  \f^{-48} & \!\!\! \simeq \!\!\! & 1 + 48 e^{-2 \pi \tii} \cos (2 \pi \ti) 
        + 600 e^{-4 \pi \tii} \cos (4 \pi \ti) + 24^{2} e^{-4 \pi \tii}, \no \\
 \no \\
  C_{1} ( \Rb , \tau ) & \!\!\! \simeq \!\!\! & ( \pt )^{1/2}
       [1 + 2 e^{- \pi \frac{\tii}{\Rb^{2}}} + 2 e^{- \pi \tii \Rb^{2}}
          + 4 e^{- \pi (\Rb^{2} + \frac{1}{\Rb^{2}})} \cos (2 \pi \ti)]. \no
\EA
Substituting these into (\ref{eq:average,mw}) and integrating about $\ti$, we obtain
\BA
  \mw
   & \!\!\! \simeq \!\!\! & - \frac{2400}{( \pa )^{\frac{1}{2}}}
   \int_{0}^{\infty} d \tii \tii^{-3/2}
   e^{- \pi (\Rb^{2} + \frac{1}{\Rb^{2}} -2)} \no \\
   && \times \left[ \TrB \right] \no
\label{eq:mwteiontotyu}
\EA
as the main part for large $\tii$. Changing the integration variable $\tii$ to $t = \frac{1}{\tii}$ and using the formula
\BE
  \int_{0}^{\infty} \frac{e^{-ax} e^{- \frac{b^{2}}{x}}}{\sqrt{x}}
    = \sqrt{\frac{\pi}{a}} e^{-2b \sqrt{a}}, \no
\label{eq:sekibunkousiki}
\EE
one obtains
\BA
  \mw
   & \!\!\! \simeq \!\!\! & 2400 \sqrt{2 \pi} \sum_{r=1}^{\infty}
   \exp \left[- \left( \frac{1}{\ad} (\Rb^{2}
   + \frac{1}{\Rb^{2}} -2) \right)^{1/2} r \B \right] \no \\
   & \!\!\! \simeq \!\!\! & 2400 \sqrt{2 \pi}
   \exp \left[- \left( \frac{1}{\ad} (\Rb^{2}
   + \frac{1}{\Rb^{2}} -2) \right)^{1/2} \B \right].
\EA
In the second equality, because $\Rb^{2} + \frac{1}{\Rb^{2}} -2 > 0$, we have kept the leading term in $\B \rightarrow \infty$ limit. In the same way, we can compute $\one$ as
\BA
  \one
  & \!\!\! \simeq \!\!\! & 24^{2} \times 50 \sqrt{2 \pi}
  \exp \left[ - \frac{\B}{{\ad}^{1/2} \Rb} \right] \no \\
  && + 24^{2} \times 50 \sqrt{2 \pi}
  \exp \left[ - \frac{\B \Rb}{{\ad}^{1/2}} \right] \no \\
  && + 2400 \sqrt{2 \pi}
  \exp \left[ - \frac{\B}{{\ad}^{1/2}}
  (\Rb^{2} + \frac{1}{\Rb^{2}})^{1/2} \right],
\label{eq:bosononeLT}
\EA
where, we have dropped the zero energy mode, namely that of $N = \Nt = 1$ and $m = w = 0$, which makes no sense physically and gives a divergent contribution. Each term of (\ref{eq:bosononeLT}) makes a major contribution for $\Rb > \sqrt{2}, \Rb < \frac{1}{\sqrt{2}}$ and $\frac{1}{\sqrt{2}} < \Rb < \sqrt{2}$, respectively. The results are summarized as follows:
\begin{enumerate}
\item \( \Rb > \sqrt{2} \)
\BE
  \frac{\mw}{\one} \simeq
  12 \exp \left[ - \frac{\B}{{\ad}^{1/2}} \left( (\Rb^{2} 
  + \frac{1}{\Rb^{2}})^{1/2} - \frac{1}{\Rb} \right) \right],
\label{eq:lRlowTbosmw}
\EE
\item \( \Rb < \frac{1}{\sqrt{2}} \)
\BE
  \frac{\mw}{\one} \simeq
  12 \exp \left[ - \frac{\B}{{\ad}^{1/2}} \left( (\Rb^{2} 
  + \frac{1}{\Rb^{2}})^{1/2} - \Rb \right) \right],
\EE
\item
\( \frac{1}{\sqrt{2}} < \Rb < \sqrt{2} \)
\BE
  \frac{\mw}{\one} \simeq 1.
\label{eq:averagemwone}
\EE
\end{enumerate}
The result of (\ref{eq:averagemwone}) implies that all strings fall into states with either $N = 1, \Nt = 0, m_{i} = \pm 1, w_{i} = \pm 1$ and other modes $0$, or $N = 0, \Nt = 1, m_{i} = \pm 1, w_{i} = \mp 1$ and other modes $0$. In all cases the momentum-winding correlation goes to $0$ or $1$ as $\B \rightarrow \infty$.

Next, we compute the momentum-winding correlation of superstring theory in 10 dimensions. Adding a fermionic part, we obtain the free energy of a superstring on $T^{9}$ in proper time form \cite{Pol}
\BA
  \FS ( \B ) = - \frac{1}{ ( \pa )^{\frac{1}{2}} } \int_{0}^{\infty} 
    \frac{d \tii }{ \pt } \int_{-1/2}^{1/2} d \ti ( \pt )^{-5}
    \thiv (C_{1})^{9} \no \\
    \times \left[ \TS \right],
\label{eq:superfreecom}
\EA
where $\vartheta$ denotes the Jacobi theta function. In consideration of a fermionic string gas together with a bosonic one, the average of $\left\langle \left( \sum_{i=1}^{9} ( m_{i} w_{i} )^{2} \right) \right\rangle$ is given by
\BA
  \mws = - \frac{1}{ ( \pa )^{\frac{1}{2}} } \int_{0}^{\infty} 
    \frac{d \tii }{ \pt } \int_{-1/2}^{1/2} d \ti ( \pt )^{-5} \thiv \no \\
    \times \left[ \sum_{r=1}^{ \infty } (1-(-1)^{r}) (-r \beta) 
    \exp \left( - \frac{r^{2} \beta^{2}}{4 \pi \ad \tii } \right) \right]
    (C_{1})^{8} \frac{9}{(2 \pi i)^{2}} \frac{\p^{2}}{\p \ti^{2}}(C_{1}).
\EA
Similarly, the average of the number of strings is given by
\BA
  \one = - \frac{1}{ ( \pa )^{\frac{1}{2}} } \int_{0}^{\infty} 
    \frac{d \tii }{ \pt } \int_{-1/2}^{1/2} d \ti ( \pt )^{-5}
    \thiv (C_{1})^{9} \no \\
    \times \left[ \sum_{r=1}^{ \infty } (1-(-1)^{r}) (-r \beta) 
    \exp \left( - \frac{r^{2} \beta^{2}}{4 \pi \ad \tii } \right) \right],
\EA
Then making use of expansion
\BE
  \thiv \simeq 1 + 32 e^{-2 \pi \tii} \cos (2 \pi \ti), \no
\EE
we can calculate $\smw$ and $\one$ in the same way as the bosonic string
\BA
  \mws & \!\!\! \simeq \!\!\! & 128 \times 9 \sqrt{2 \pi}
   \exp \left[ - \left(\frac{1}{\ad}
    (\Rb^{2} + \frac{1}{\Rb^{2}} -2) \right)^{1/2} \B \right], \\
  \one & \!\!\! \simeq \!\!\! & 36 \sqrt{2 \pi} \left[ \exp
     \left( - \frac{\B}{{\ad}^{1/2} \Rb} \right)
        + \exp \left( - \frac{\B \Rb}{{\ad}^{1/2}} \right) \right].
\EA
The results are
\begin{enumerate}
\item \( \Rb > 1 \)
\BE
  \frac{\mws}{\one} \simeq
  32 \exp \left[ - \frac{\B}{{\ad}^{1/2}} \left( (\Rb^{2} 
  + \frac{1}{\Rb^{2}})^{1/2} - \frac{1}{\Rb} \right) \right],
\label{eq:lRlowTsupmw}
\EE
\item \( \Rb < 1 \)
\BE
  \frac{\mws}{\one} \simeq
  32 \exp \left[ - \frac{\B}{{\ad}^{1/2}} \left( (\Rb^{2} 
  + \frac{1}{\Rb^{2}})^{1/2} - \Rb \right) \right].
\label{eq:sRlowTsupmw}
\EE
\end{enumerate}
Therefore, even in the superstring case, the momentum-winding correlation goes to zero as $\B \rightarrow \infty$. These results lead to the conclusion that in the low temperature limit, the momentum-winding correlations of both bosonic strings and superstrings are very small. Therefore, there is no possibility of the existence of sharp wave packets in both $x$-space and $\xt$-space. This implies no possibility of communicating information from $\xt$-space to $x$-space in the low temperature.

\section{The momentum-winding correlation near the \\ Hagedorn temperature}
\label{sec:high}

When the radius of the torus universe is a minimum ($\Rb = 1$), it is natural to think that the energy density and temperature of strings are extremely high. As Sathiapalan, \cite{Sa} Kogan, \cite{Ko} Atick and Witten \cite{AW} have pointed out, if the energy density is sufficiently large, coupling becomes strong and a phase transition may occur in this region. However, in so far as we do not know string thermodynamics in the strong coupling region yet, let us assume the coupling of strings is extremely small (the string length is much longer than the Plank length) in this section. A cosmological model with small coupling constant will be considered in the next section.

In the weak coupling assumption, we can treat strings as an ideal gas, and it is well-known that temperature of a string gas has an upper limit --- the Hagedorn temperature $T_{H}$, since the degeneracies of oscillation modes grow exponentially with the eigenvalues of operators $N$ and $\Nt$ . Near $T_{H}$, we cannot believe the canonical ensemble method works as in the low temperature case, because energy fluctuations will be very large in this method. \cite{Efura} In this section, therefore, we discuss the momentum-winding correlation near the Hagedorn temperature by using the more fundamental microcanonical ensemble method. 

First, we consider bosonic strings on a 25-dimensional torus $T^{25}$. As the expectation value of winding number is evaluated in Ref.\cite{Tan3}, we compute the momentum-winding correlation in the same way. Densities of states of oscillation modes $N$ and $\Nt$ for large eigenvalues are \cite{GSW}
\BA
  d_{N} & \!\!\! \simeq \!\!\! &
    \frac{1}{\sqrt{2}} N^{-27/4} e^{4 \pi \sqrt{N}}, \no \\
  d_{\Nt} \!\!\! & \simeq \!\!\! &
    \frac{1}{\sqrt{2}} \Nt^{-27/4} e^{4 \pi \sqrt{\Nt}} \no .
\EA
From these equations, we see that the number of oscillation modes increases exponentially at high energy so that in high temperature limit, most of the energy comes from the oscillation modes. This is the origin of the Hagedorn temperature. For fixed quantum numbers $m$ and $w$, the density of a single string state $f( \e, m, w)$ is given by
\BA
  f( \e ,m,w) & \!\!\! = \!\!\! & d_{N( \e ,m,w)} d_{\Nt( \e ,m,w)} 
   \frac{\p}{\p \e} N( \e ,m,w) \no \\
    & \!\!\! \simeq \!\!\! & \frac{1}{4} \ad \e (N \Nt)^{-27/4}
     e^{4 \pi \left( \sqrt{N} + \sqrt{\Nt} \right)}. \no
\label{eq:femw1}
\EA
For the given single string energy $\e$ and momenta $p_{i} \equiv {\ad}^{-1/2} ( m_{i} / \Rb - w_{i} \Rb )$, $\tilde{p}_{i} \equiv {\ad}^{-1/2} ( m_{i} / \Rb + w_{i} \Rb )$, we can use the approximation $p_{i} / \e$, $\tilde{p}_{i} / \e\ll 1$ when the masses of strings are much larger than the momentum or winding energy in the high temperature limit. In this approximation, $f( \e, m, w)$ can be rewritten
\BE
  f( \e ,m,w) \simeq \frac{1}{\e} \left( \frac{2}{{\ad}^{1/2} \e} \right)^{25}
   e^{\BH \e} e^{- A_{i} m_{i}^{2} - B_{i} w_{i}^{2}},
\label{eq:femw2}
\EE
where $A_{i}$ and $B_{i}$ are
\BA
  A_{i} &=& \frac{2 \pi}{{\ad}^{1/2} \Rb_{i}^{2} \e }, \no\\
  B_{i} &=& \frac{2 \pi \Rb_{i}^{2}}{{\ad}^{1/2} \e }, \no
\EA
and $\BH = 4 \pi {\ad}^{1/2}$ is the inverse Hagedorn temperature. Summing over $m$ and $w$, we obtain the single-string degeneracy $f( \e )$,
\BA
  f( \e ) & \!\!\! = \!\!\! & \sum_{m,w} f( \e ,m,w) \no \\
   & \!\!\! = \!\!\! & \frac{1}{\e}
    \left( \frac{2}{{\ad}^{1/2} \e} \right)^{25}
     e^{\BH \e} \sum_{m,w} e^{- A_{i} m_{i}^{2} - B_{i} w_{i}^{2}} .
\label{eq:fe1}
\EA
From this point, we consider the case in which all torus radii satisfy $\Rb \sim 1$. When $\e$ satisfies the conditions $\frac{1}{{\ad}^{1/2} \Rb^{2}} \ll \e$ and $\frac{\Rb^{2}}{{\ad}^{1/2}} \ll \e$, namely for $\frac{1}{{\ad}^{1/2}} \ll \e$, we can approximate the summation over $m$ and $w$ by the Gaussian integral
\BA
  f( \e ) & \simeq & \frac{1}{\e} \left( \frac{2}{ {\ad}^{1/2} \e} \right)^{25}
   e^{\BH \e} \prod_{i=1}^{25}
    \left( \frac{\pi^{2}}{A_{i} B_{i}} \right)^{1/2} \no \\
    & = & \frac{1}{\e} e^{\BH \e}.
\label{eq:fe2}
\EA
On the other hand, the number of states is given by
\BE
  \O (E) \simeq \BH \exp \left[ \BH E + \xi (\BH) \right],
\label{eq:compactmicro}
\EE
where $E$ is the total energy of the string gas, and $\xi$ is an analytic function of $\BH$, as calculated by complex temperature method in Ref.\cite{Tan2}. The average of the square of the momentum $m$ times the winding number $w$ is calculated from the formula
\BE
  \mw = \int_{\L_{1}}^{E} d \e \frac{1}{\O (E)}
    \left[ \sum_{i=1}^{25} \sum_{m_{i},w_{i}} (m_{i} w_{i})^{2}
      f( \e ,m,w) \right]
       \O (E- \e ),
\label{eq:mwaverage}
\EE
where $E$ is the total energy of strings and $\L_{1}$ is the cutoff which is needed for the approximation in (\ref{eq:fe2}), and is chosen as the larger value among $\frac{1}{{\ad}^{1/2} \Rb^{2}}$ and $\frac{\Rb^{2}}{{\ad}^{1/2}}$. The sum in the square brackets is calculated from (\ref{eq:fe2}) by replacing $(mw)^{2}$ with derivatives about $A$ and $B$, i.e.,
\BA
  \sum_{i=1}^{25} \sum_{m_{i},w_{i}} (m_{i} w_{i})^{2} f( \e ,m,w) 
   & \!\!\! = \!\!\! &
         \sum_{i=1}^{25} \frac{\p}{\p A_{i}} \frac{\p}{\p B_{i}}
       \sum_{m_{i},w_{i}} f( \e ,m,w) \no \\
   & \!\!\! \simeq \!\!\!\! & 
         \frac{25^{2}}{4} \frac{\ad}{(2 \pi)^{2}} \e e^{\BH \e}.
\label{eq:ABderivative}
\EA
Substituting this into (\ref{eq:mwaverage}), we obtain
\BE
  \mw \simeq \frac{25^{2} \ad}{32 \pi^{2}} E^{2}.
\EE
This must be divided by the number of strings $\one$, which is
\BA
  \one & \!\!\! \simeq \!\!\! & \int_{\L_{0}}^{E} d \e
        \frac{ f ( \e ) \O (E- \e )}{\O (E)} \no \\
    & \!\!\! \simeq \!\!\! & \ln \frac{E}{\L_{0}},
\EA
where $\L_{0} = \frac{1}{{\ad}^{1/2}}$ is the cutoff which is necessary for oscillation mode dominance approximation. Therefore, the momentum-winding correlation is
\BA
  \frac{{ \displaystyle \mw }}{\one} & \simeq & \frac{25^{2} \ad}{32 \pi^{2}} 
    \frac{E^{2}}{\ln {\displaystyle \frac{E}{\L_{0}}}} \no \\
  & \simeq & \frac{25^{2} \ad}{32 \pi^{2}} 
    \frac{C^{2}}{(\B - \BH ) \ln \left[ 
      {\displaystyle \frac{C}{(\B - \BH )^{1/2} \L_{0}}} \right] } \no \\
  & \simeq & \frac{25^{2} \ad}{32 \pi^{2}} 
    \frac{S^{2}}{\BH^{2} \ln {\displaystyle \frac{S}{\BH \L_{0}}}},
\label{eq:mwuncertain}
\EA
where $S$ represents for the total entropy. In deriving (\ref{eq:mwuncertain}), we have used the fact that the inverse temperature and total energy of the string gas in the Bolzmann approximation are given by \cite{BVcos}
\BA
  E & \!\!\! \simeq \!\!\! &
        \left[ \frac{b \L_{1}}{a( \B - \BH )} \right]^{1/2} 
   = C( \B - \BH )^{-1/2}, \no
\label{eq:EB} \\
  \B - \BH & \!\!\! \simeq \!\!\! & \frac{b \L_{1} \BH^{2}}{aS^{2}}
   = C^{2} \frac{\BH^{2}}{S^{2}}, \no
\label{eq:BS}
\EA
where, $a = 0.56$, $b = 0.29$ and $C \equiv ( \frac{b \L_{1}}{a} )^{1/2}$. We can see from (\ref{eq:mwuncertain}) that if $E$ is much larger than $\L_{0}$ (that is, if $\B$ is very close to $\BH$ and $S$ is much larger than $1$) then the momentum-winding correlation is very large.

Here, we consider the uncertainty principle between the two coordinates $x$ and $\xt$ of strings. The average of the square of momentum number about a single string can be calculated by applying the derivative $\p / \p A_{i}$ to $f$, just as in (\ref{eq:ABderivative}), and we obtain
\BE
  ( \triangle m )^{2} = \frac{{ \displaystyle \left\langle \left( \sum_{i=1}^{25}
    m_{i}^{2} \right) \right\rangle }}{\one}
    \simeq \frac{25 {\ad}^{1/2} \Rb^{2}}{4 \pi} 
    \frac{E}{\ln {\displaystyle \frac{E}{\L_{0}}}} \no .
\EE
For the winding number, the fluctuation can be calculated similarly by applying $\p / \p B_{i}$ to $f$. These results are combined to obtain the uncertainty relation between the momentum and the winding number as
\BE
  \triangle m \cdot \triangle w  
    \simeq \frac{25 {\ad}^{1/2}}{4 \pi} 
    \frac{E}{\ln {\displaystyle \frac{E}{\L_{0}}}},
\EE
From the relation (\ref{eq:xxuncertainty}), the uncertainty of the two coordinates $x$ and $\xt$ is given by
\BE
  \triangle x \cdot \triangle \xt 
    \simeq \frac{4 \pi \hbar^2}{25 {\ad}^{1/2}}
            \frac{\ln {\displaystyle \frac{E}{\L_{0}}}}{E}.
\EE
This means that if $E$ is very large, there can be a wave packet sharp in both $x$- and $\xt$-space.

Next, we consider the case in which the torus radii in $\db$ dimensions are much larger than $1$ and the temperature is kept near the Hagedorn temperature. In the region $\L_{0} < \e < \L_{1}$, we can approximate the sum of the winding modes $w$ in $\db$ dimensions in (\ref{eq:fe1}) by $1$, i.e., $\sum_{i=1}^{\db} \sum_{w_{i}} e^{- B_{i} w_{i}^{2}} =1$. Then we obtain single string degeneracy:
\BE
  f( \e ) = 2^{\db /2} \Rb^{\db}
     \frac{\exp (\BH \e)}{{\ad}^{\db /4} \e^{\db / 2 +1}}.
\EE
For the case $\e > \L_{1}$, we can approximate all the sums in (\ref{eq:fe1}) by a Gaussian integral. We then obtain (\ref{eq:fe2}). According to Refs.\cite{Tan2} and \cite{Tan3}, number of states in this case is
\BE
  \O_{0} (E, \Rb) = \BH \exp \left[ \BH E + \Rb^{\db} ( \xi_{0} +a)
        + O(\Rb^{\db -1}, \ln \Rb) \right],
\label{eq:largeRmicro}
\EE
when the total energy $E$ is larger than $\L_{1}$, and from these equations, the momentum-winding correlation turns out to be
\BE
  \frac{{ \displaystyle \left\langle \left( \sum_{i=1}^{\db}
   ( m_{i} w_{i} )^{2} \right)
    \right\rangle }}{\one} \simeq \frac{\db^{2} \ad}{32 \pi^{2}} 
     \frac{E^{2} - \L_{1}^{2}}{{\displaystyle \frac{c' \Rb^{\db}}{\db /2 +2}
      \left( \frac{1}{\L_{0}^{\db /2+2}} - \frac{1}{\L_{1}^{\db /2+2}} \right)
       + \ln \frac{E}{\L_{1}}}}.
\label{eq:largeRhighTsupermw}
\EE
This result indicates that even in the case of large $\Rb$, the momentum-winding correlation is very large if $E$ is sufficiently large.

As for the superstring, the momentum-winding correlation can be obtained in the same way as the bosonic string. The difference comes from the degeneracies of oscillating modes, \cite{GSW}
\BA
  d_{N} & \!\!\! \simeq \!\!\! & (2N)^{-11/4} e^{\pi \sqrt{8N}}, \no \\
  d_{\Nt} & \!\!\! \simeq \!\!\! & (2 \Nt)^{-11/4} e^{\pi \sqrt{8 \Nt}}, \no
\EA
and the inverse Hagedorn temperature is $\BH = 2 \pi \sqrt{2 \ad}$. The result is
\BA
  \frac{{ \displaystyle \mws }}{\one} & \!\!\! \simeq \!\!\! &
   \frac{9^{2} \ad}{16 \pi^{2}} 
    \frac{E^{2}}{\ln {\displaystyle \frac{E}{\L_{0}}}} \no \\
  & \!\!\! \simeq \!\!\! & \frac{9^{2} \ad}{16 \pi^{2}} 
     \frac{C^{2}}{(\B - \BH ) \ln \left[ 
      {\displaystyle \frac{C}{(\B - \BH )^{1/2} \L_{0}}} \right] } \no \\
  & \!\!\! \simeq \!\!\! & \frac{9^{2} \ad}{16 \pi^{2}} 
     \frac{S^{2}}{\BH^{2} \ln {\displaystyle \frac{S}{\BH \L_{0}}}},
\label{eq:supermwuncertain}
\EA
which is the same as that of the bosonic string, except for a numerical factor. Therefore, even in the superstring case, the momentum-winding correlation is very large when the total energy of the string gas is very large. These results lead to the conclusion that near the Hagedorn temperature, the momentum-winding correlations of both bosonic strings and superstrings are very large. Therefore, there is a possibility of the existence of sharp wave packets in both $x$-space and $\xt$-space. This implies the possibility of communicating information from $\xt$-space to $x$-space.

\section{Inflationary cosmological model}
\label{sec:inflation}

\begin{figure}
\begin{center}
$${\epsfxsize=13.5 truecm \epsfbox{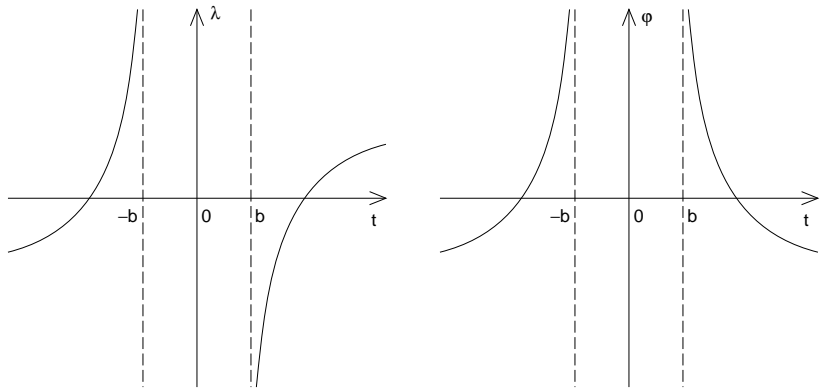}}$$
\caption{The time development of scale factors and shifted dilatons in the Tseytlin model.}
\label{fig:Tsey_graph}
\end{center}
\end{figure}

In the previous section, we saw that there is a possibility of exchanging information between $x$-space and $\xt$-space near the Hagedorn temperature. In order to investigate whether we can observe any trace of the $\xt$ universe in our $x$ universe, we must compare the expansion rate of the universe to the interaction rate of strings, and we must examine whether any information of the previous world has been kept today. Let us first consider the time evolution of the background metric of the universe. In the case that the torus universe is filled with string gas near the Hagedorn temperature, the time dependence of the scale factor has been solved in the low energy effective action by Tseytlin. \cite{Tseycos} The low-energy effective action is
\BE
  S_{26} = - \frac{1}{2} \int d^{26} x \sqrt{g} e^{-2 \phi}
      ( {\cal R} - 4 \nabla_{\mu} \phi \nabla^{\mu} \phi ),
\label{eq:effaction}
\EE
where $g$ is the determinant of the metric, $\phi$ the dilaton and ${\cal R}$ the scalar curvature. Here, we ignore the antisymmetric tensor. Then, we take the string gas effect into consideration with the matter action
\BE
  S_{m} = \frac{1}{2} \int d^{26} x \sqrt{g} T,
\EE
where $T$ is the trace of the energy-momentum tensor. We must note that we cannot trust this action in a high curvature region, but we ignore this effect at this point. For simplicity, we assume that the universe is homogeneous and isotropic and that the radii of the 22 directions of the torus are kept compactified and stable, and these effects can be disregarded. Then we adopt the line element of the remaining 3+1 dimensions as the flat Robertson-Walker type with a common scale factor in 3 spatial directions:
\BE
  ds^{2} = - dt^{2} + \sum_{i=1}^{3} R^{2} (t) dx_{i}^{2}. \no
\EE
The energy momentum tensor is assumed to be that of a perfect fluid with energy density $\rho$ and pressure $P$:
\BE
  T_{\mu}^{\nu} = diag ( - \rho (t) , P(t) \delta_{i}^{j} ).
\EE
The radius of the torus is chosen as the scale factor of the universe. As we can see from the number of states (\ref{eq:compactmicro}) or (\ref{eq:largeRmicro}), the equation of state near the Hagedorn temperature is given by $P \simeq 0$ since the energy of the strings is dominated by oscillation modes and the kinetic energy is very small. Using these assumptions, we can solve the equations derived from the low energy effective action, as was done by Tseytlin, \cite{Tseycos}
\BA
  \vp & \!\!\! = \!\!\! & \vp_{0} - \ln \mid t^{2} - b^{2} \mid, \\
  R & \!\!\! = \!\!\! & R_{0} \left| \frac{t-b}{t+b} \right|^{1/ \sqrt{3}}.
\label{eq:Rdevelop}
\EA
Here $R_{0}$ is the radius of the torus in the $t \rightarrow - \infty$ limit and $\vp$ is a shifted dilaton $\vp \equiv 2 \phi - 3 \ln \Rb$ which is invariant under the duality transformation. In Fig.\ref{fig:Tsey_graph} to display the symmetry of solution, we use $\l$ instead of $R$, which is defined by the relation $\Rb (t) = \frac{R (t)}{\sqrt{\ad}} \equiv e^{\l (t)}$.

As we can see from Fig.\ref{fig:Tsey_graph}, this solution has two regions, the inflationary one in $t<-b$ and deflationary one in $t>b$. In the inflationary region, since both $\l$ and $\vp$ are increasing, $\phi$ is also increasing, as seen from the definition of $\vp$. In order to be consistent with the weak coupling approximation in \S \ref{sec:high}, we wish to choose the inflationary solution, because the coupling is very small when $\Rb \sim 1$. However, there is the graceful exit problem in connecting this inflation and the present deflation, i.e., matter dominating universe. Veneziano et al. have proposed a pre-big bang string cosmology model, \cite{GVcos} in which the target space spreads out infinitely unlike the our torus model. There, both scale factors and string sizes extend consistently in the low energy effective action method. They attempted to connect inflation and deflation phases by introducing a dilaton potential which is expected to appear as a non-perturbative effect. However, for this connection of the two phases no-go theorem was proved by Kaloper, Madden and Olive \cite{nogo} (see also Ref.\cite{grace}). Therefore, we cannot connect those two solutions in the low-energy effective action with dilaton potential when the equation of the state of string matter is given by $P= \gamma \rho$ with $-1< \gamma$. Although, this no-go theorem is applicable to our model, where $\gamma = 0$, we simply assume that the connection of the two phases is possible for some unspecified reason in the following arguments.

Second, we wish to see the interaction rate of strings. However, we do not know the high energy interaction of strings precisely enough. We would like to focus on the gravitational effect to metric perturbation. We suppose that there were energy fluctuations of strings in $x$-space created by the momentum-winding transition, and consider whether this information has been kept to the present day. If the expansion rate of the universe is larger than the gravitational interaction rate, then the metric fluctuation is frozen during the inflation. In general, however, inflation models are introduced for solving smoothness problems, that is, we cannot observe metric fluctuations created before the inflation at least within our photon horizon today. In fact, the cosmic no hair theorem has been discussed by many people. Therefore, it seems that the fluctuations created by the momentum-winding uncertainty is not kept today within the inflation model. Of course, the string cosmology model is different from the usual inflation models at the point that the gravitational coupling changes in time, and there is little hope that the energy density fluctuation is amplified by the Jeans instability. However, this is shown not to be the case below. The Jeans length is defined by
\BE
  L_{J}^{2} = \frac{1}{G \rho},
\EE
where $G$ is the gravitational coupling and $\rho$ is the energy density of strings. In our case, $G=e^{2 \phi}$ and $\rho = E R^{-3}$, so that
\BE
  L_{J} = \left( \frac{ | t^{2} - b^{2} | }
     { {\ad}^{1/2} E e^{ \varphi_{0}}} \right)^{1/2}.
\EE
As time passes, in spite of the decrease of the energy density $\rho$, $L_{J}$ continues to decrease because the coupling $G$ increases more rapidly. From this it seems that $L_{J}$ will be smaller than the string extension $\triangle x$, and the fluctuation of energy density will be amplified by the Jeans instability. As the universe is in the inflation phase, however, the event horizon appears. If $L_{J}$ is larger than this horizon size, the Jeans instability is ineffective, since the gravitational force does not reach beyond the horizon. The event horizon is defined by
\BE
  L_{H} = R(t) \int_{t}^{t_{max}} \frac{1}{R(t')} dt',
\label{eq:LHRt}
\EE
where, in our case, $t_{max}$ is $-b$. If (\ref{eq:Rdevelop}) is substituted into (\ref{eq:LHRt}), $L_{H}$ is represented in the form of an infinite sum,
\BE
  L_{H} = - (t+b) - \frac{2b}{\sqrt{3}} \sum_{n=1}^{\infty}
    \frac{1}{1/ \sqrt{3} + n} \left( \frac{t+b}{t-b} \right)^{n},
\label{eq:horizon}
\EE
so that $L_{H}$ also shrinks as $t$ approaches $-b$. Since the infinite sum in $L_{H}$ is complicated, we pick up the first term to compare $L_{J}$ and $L_{H}$, and consider the case for $L_{J} < - (t+b)$. This condition implies $R < 2^{1/ {2 \sqrt{3}}} R_{0}$, so that the horizon scale becomes smaller than the Jeans length long before the universe expands rapidly ($t \ll -b$). If we include the infinite sum in (\ref{eq:horizon}) the upper bound is lower than this. Therefore, there is no possibility of the amplification of the energy density fluctuation as a result of the Jeans instability. This implies that the metric perturbation coming from momentum-winding transition is frozen during the inflation phase, and we cannot observe this perturbation if the universe has been expanded by inflation, and thereby the wavelength of fluctuations become longer than the particle horizon at present.

\section{Summary}
\label{sec:conclusion}

In this paper we have analyzed the momentum-winding transitions in the Brandenberger-Vafa model by taking account of the momentum-winding correlations. As we have seen in \S \S \ref{sec:low} and \ref{sec:high}, the momentum-winding correlation is very small in the low temperature limit, while very large near the Hagedorn temperature. Therefore, in the low temperature limit, there is no possibility of the existence of sharp wave packets in both $x$-space and $\xt$-space, which implies no possibility of exchanging information from $\xt$- to $x$-space. Contrastingly, near the Hagedorn temperature, there is a possibility of the existence of sharp wave packets in both $x$- and $\xt$-space, and of exchanging information from $\xt$- to $x$-space. Then we considered the inflational string cosmological model based on the calculation of Tseytlin using the low energy effective action method. Investigating whether the amplification of energy density fluctuations occurs through the Jeans instability, we have concluded that there is no possibility of this amplification in this model. This means that if the period of inflational phase is sufficiently long, we cannot observe this fluctuation. In any case, we must resolve the no-go theorem and construct a model of a universe for the whole time region in order to discuss whether the fluctuations would remain in our horizon today.

\section*{Acknowledgements}

The authors thank K. Ohta for helpful suggestions for calculations. This work is supported in part by the Grant-in-Aid for Scientific Research from the Ministry of Education, Science and Culture, No. 06640398 and No. 08650457. The work is also supported in part by the Japan-Former Soviet Union Scientists Collaboration Program.

\end{document}